\documentclass[twocolumn,aps,prl] {revtex4} %% 2-spaltig
\usepackage{graphicx}
\begin{document}
\pagestyle{empty} 
\title{Rubber friction on smooth surfaces}
\author{  
B.N.J. Persson$^1$ and A.I. Volokitin$^{1,2}$} 
\affiliation{$^1$IFF, FZ-J\"ulich, 52425 J\"ulich, Germany}
\affiliation{$^2$Samara State Technical University, 443100 Samara, Russia}

\begin{abstract}
We study the sliding friction for viscoelastic solids, e.g., rubber,
on hard flat substrate surfaces. We consider first the fluctuating shear stress
inside a viscoelastic solid which result from the thermal motion of the atoms 
or molecules in the solid. At the nanoscale the thermal fluctuations are very strong
and give rise to stress fluctuations in the MPa-range, which is similar to the depinning stresses
which typically occur at solid-rubber interfaces, indicating the crucial importance of
thermal fluctuations for rubber friction on smooth surfaces. We develop a detailed model which
takes into account the influence of thermal fluctuations on the depinning of small 
contact patches (stress domains) at the rubber-substrate interface. The theory predict
that the velocity dependence of the macroscopic shear stress has a bell-shaped form, and that the
low-velocity side exhibit the same temperature dependence as the bulk viscoelastic modulus,
in qualitative agreement with experimental data. Finally, we discuss the influence
of small-amplitude substrate roughness on rubber sliding friction. 
\end{abstract}
\maketitle

%%%%%%%%%%%%%% main text %%%%%%%%%%%%%%%%
%\begin{multicols}{2}

\vskip 0.5cm

{\bf 1. Introduction}

The friction between rubber and 
smooth substrate surfaces is a topic of extreme practical importance,
e.g.,
for wiper blades (in particular on hydrophobic glass),  rubber O-ring seals, and in the
contact region between the tire-rubber and the steel rim on a wheel\cite{Moore}.

\begin{figure}
\includegraphics[width=0.45\textwidth,angle=0]{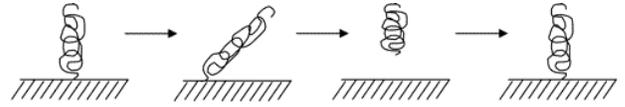}
\caption{\label{Schall}
The classical description of a polymer chain in contact with a lateral moving countersurface. The chain
stretches, detaches, relaxes, and reattaches to the surface to repeat the cycle.
Adapted from Ref. \cite{Cha}.}
\end{figure}

When a rubber block is sliding on a very rough substrate, such as a tire on a road surface,
the friction is almost entirely due to the energy dissipation in the bulk of the rubber as a result of
the fluctuating (in time and space) viscoelastic deformations 
of the rubber by the substrate asperities\cite{P8,PV,Heinrich}.
This mechanism becomes unimportant when the substrate is very smooth. In the limiting case of a
perfectly smooth substrate, the friction is instead due to local stick-slip events at the sliding interface.
Schallamach\cite{Schall1} 
has proposed a molecular mechanism for the local stick slip, see Fig. \ref{Schall}, where
rubber polymer chains at the interface attach to the moving countersurface, 
stretches, detaches, relaxes, and reattaches to the surface to repeat the cycle
(similar models have been studied in Ref. \cite{Cher,Urbach}). During each cycle,
the elastic energy stored in the polymer chain is dissipated as heat during the (rapid) detachment
and relaxation phase, and this is assumed to be the origin of the (macroscopic) friction. However,
in our opinion this picture cannot be fully correct. First, for physisorption systems
the energy barriers for (vertical) detachment are usually much
higher than the energy barriers for lateral sliding\cite{Woll}, 
so one would not expect any detachment to occur.
Secondly, with respect to the stresses rubber materials are usually exposed to, 
rubber is nearly incompressible and it is not easy to imagine how single molecules 
strongly confined at the interface are able to switch between an elongated (stretched) 
state and a relaxed (curled up) state as indicated in the figure. Furthermore, the 
sliding friction tend to exhibit the same temperature dependence as the bulk
rubber viscoelastic modulus $E(\omega)$ as described by the Williams-Landel-Ferry (WLF)\cite{WLF}
shift factor $a_T$. This indicate the crucial role of the bulk rubber in the friction process.

\begin{figure}
\includegraphics[width=0.35\textwidth,angle=0]{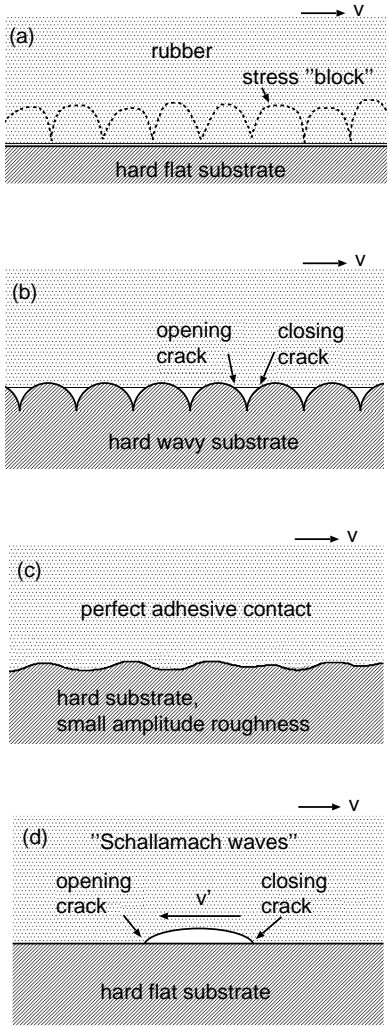}
\caption{\label{four}
For different processes where the adhesive rubber-substrate interaction contribute
to rubber friction.
}
\end{figure}

We believe that the local stick-slip processes, which must occur at the sliding interface,
involve relative larger rubber volume elements,
always in adhesive contact
with the substrate.
That is, during sliding small patches (with a diameter
of order $D\sim 10-100 \ {\rm nm}$) or {\it stress domains}\cite{Persson1}
of rubber at the sliding interface
perform stick-slip motion: during stick the shear stress at the interface increases continuously with time
until the local shear stress reaches a critical {\it depinning} stress $\sigma_{\rm c}$, after which 
a rapid local slip occur, but with the rubber patch in continuous 
adhesive contact with the substrate. 
During the local slip the elastic deformation 
energy stored in the rubber during the loading phase
will be dissipated party inside the rubber (in a volume of order $D^3$)
and partly at the interface. 
The deformation field in the vicinity of a stress domain of area $\sim D^2$ will extend a distance
$\sim D$ into the rubber block; we denote this basic unit 
(volume $\sim D^3$) as a {\it stress block} [see volume elements surrounded by dashed lines in 
Fig. \ref{four}(a)]. 

Figure \ref{four} illustrate three other mechanisms of rubber friction, which all depend on
the rubber-substrate adhesional interaction. Fig. \ref{four}(b) illustrate a case where a rubber
block is sliding on a smooth wavy substrate. Here we assume only roughness (waviness) on a single
length scale, i.e., the substrate bumps have no roughness on length scales smaller
than the lateral size of the bumps. In this case, if the adhesional interaction
(and the external applied normal load) is unable to bring the rubber into perfect contact with the substrate,
(as in the figure), during sliding opening and closing cracks will occur at the exit and the front of
each asperity contact region. In general, negligible bulk viscoelastic energy
dissipation occur at the closing crack, while a huge energy dissipation may occur at the
opening crack\cite{Ueba,Brener}. It has been shown that in some cases the dominant contribution to the 
friction force arises from 
the energy ``dissipation'' at the 
opening cracks\cite{Ueba,Brener}. 

If the substrate has small-amplitude roughness, the adhesive rubber-substrate interaction may
(even in the absence of an external load)
lead to complete contact between the rubber and the substrate 
at the sliding interface [see Fig. \ref{four}(c)].
In this case, during sliding viscoelastic deformations will occur in the bulk of the rubber in the vicinity of
the substrate, which will contribute to the observed friction. This contribution can be calculated 
using the theory developed in Ref. \cite{P8,PerssonSS}.

Finally, for very soft rubber it has often been observed that
detachment waves propagate throughout the entire contact area, from its advancing to the
trailing edge, see Fig. \ref{four}(d). Schallamach\cite{Schall2} 
first discovered these waves at ``high'' sliding 
velocities and for (elastically) soft rubber. 
Roberts and Thomas\cite{RT,Ueba} have shown that when such instabilities occur, the frictional stress 
is mainly due to the energy dissipation at the opening crack, i.e., similar to the situation in Fig.
\ref{four}(b).

In this paper we study the rubber friction process shown in Fig. \ref{four}(a). This is
probably the most important rubber friction mechanism in most applications involving {\it very}
smooth surfaces. 
In Sec. 2 we consider the fluctuating shear stress
inside a viscoelastic solid, which result from the thermal motion of the atoms 
or molecules in the solid. At the nanoscale the thermal fluctuations are very strong
giving stress fluctuations in the MPa-range, which is similar to the depinning stresses
which typically occur at solid-rubber interfaces, illustrating the crucial importance of
thermal fluctuations for rubber friction on smooth surfaces. 
In Sec. 3 we develop a detailed model which
takes into account the influence of thermal fluctuations on the depinning of small 
contact areas (stress domains) at the rubber-substrate interface. The theory predict
that the velocity dependence of the macroscopic shear stress has a bell-shaped form
(see Sec. 4), and that the
low-velocity side exhibit the same temperature dependence as the bulk viscoelastic modulus,
in qualitative agreement with experimental data (Sec. 5). 
In Sec. 6 we discuss the role of 
the surface roughness, which exist even on very smooth
surfaces. We show that the countersurfaces used in most rubber sealing applications have such large
roughness that the friction observed during sliding at 
low velocity in many cases may be mainly due to the
substrate asperity-induced viscoelastic deformation of the rubber surface.
Sec. 7 contains the summary and conclusion.

\begin{figure}
\includegraphics[width=0.25\textwidth,angle=0]{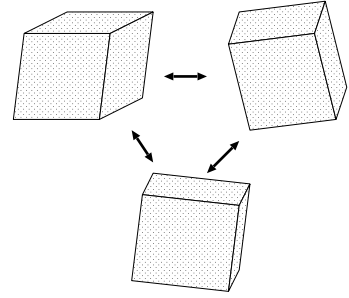}
\caption{\label{small}
The thermal motion of the atoms in a solid gives rise to shape fluctuations for a small
volume element in the solid (schematic).
}
\end{figure}

\vskip 0.5cm
{\bf 2. Brownian motion in viscoelastic solids}

A small particle in a liquid perform random motion caused by the impact of the 
surrounding liquid molecules. Similarly, a small
volume element in a viscoelastic solid perform {\it shape fluctuations} as a result of the 
thermal motion of the atoms or molecules in the solid, 
see Fig. \ref{small}. Here will will estimate the magnitude of the fluctuating shear stress 
which act on any (internal) surface in a viscoelastic solid.

Assume that ${\bf u}({\bf x},t)$ is the displacement vector in an infinite viscoelastic solid.
The equation of motion for ${\bf u}$ is
$$\rho {\partial^2 {\bf u} \over \partial t^2} = \hat \mu \nabla^2 {\bf u}+(\hat \mu + \hat \lambda)
\nabla \nabla \cdot {\bf u}+ {\bf f}\eqno(1)$$
where $\hat \mu$ and $\hat \lambda$ are time integral operators, e.g.,
$$\hat \mu G(t) = \int_{-\infty}^t dt' \ \mu (t-t') G(t'),$$
and where ${\bf f}({\bf x},t)$ is a randomly fluctuating force density
(see below).
If we define the Fourier transform
$$f_i({\bf k},\omega) = {1\over (2\pi)^4} \int d^3x dt \ f_i({\bf x},t) e^{-i({\bf k}\cdot x -\omega t)}$$
and
$$f_i({\bf x}, t) = \int d^3k d\omega \ f_i({\bf k},\omega) e^{i({\bf k}\cdot x -\omega t)}$$
we can write
$$\langle f_i({\bf k},\omega)f_j ({\bf k'},\omega')\rangle = -{k_{\rm B} T \over \pi \omega} (2\pi)^{-3}$$
$$\times \left ( {\rm Im} \mu(\omega)k^2 \delta_{ij}+{\rm Im} {\mu(\omega)\over 1-2\nu}
k_ik_j \right )$$
$$\times \delta ({\bf k}+{\bf k}') \delta (\omega + \omega')$$ 
where $\nu =\lambda/2(\mu+\lambda)$ and where
$$\mu(\omega) = \int_0^\infty dt \ \mu(t) e^{i\omega t}.$$
Let us study the fluctuating stress in the solid. The average stress $\langle \sigma_{ij}\rangle = 0$
but the average of the square of any components of $\sigma_{ij}$ will be finite. Here we consider
the magnitude of the fluctuating shear stress within some plane which we can take to be the $xy$-plane.
Thus we consider
$$\langle \sigma_\parallel^2\rangle = \langle \sigma_{zx}^2+\sigma_{zy}^2\rangle =
2\langle \sigma_{zx}^2\rangle $$
The stress tensor
$$\sigma_{ij} = \hat \mu (u_{i,j}+u_{j,i})+\hat \lambda u_{k,k}\delta_{ij}$$
Thus we get 
$$\sigma_{zx}=\hat \mu (u_{z,x}+u_{x,z})$$
and
$$\langle \sigma_\parallel^2\rangle = 
4\langle (\hat \mu u_{z,x})^2+\hat \mu u_{z,x} \hat \mu u_{x,z}\rangle $$
$$=-4 \int d^3k d^3k' d\omega d\omega' \mu(\omega) \mu(\omega')$$
$$\times ( k_x k_x'
\langle u_z({\bf k},\omega)u_z({\bf k'},\omega')\rangle$$
$$ +
k_xk_z' \langle u_z({\bf k},\omega)u_x({\bf k'},\omega')\rangle)\eqno(2)$$  
Using (1) and neglecting inertia effects gives
$${\bf u}({\bf k},\omega) = {1\over \mu(\omega)k^2 +(\mu(\omega)+\lambda(\omega)){\bf k}{\bf k}}
\cdot {\bf f}({\bf k},\omega)$$
$$={1\over \mu(\omega) k^2}\left (1-{1\over 2(1-\nu)} {{\bf k}{\bf k}\over k^2}\right )
\cdot  {\bf f}({\bf k},\omega).\eqno(3)$$
Using (2) and (3) and assuming that $\nu$ is frequency independent gives
$$\langle \sigma_\parallel^2 \rangle = 
-k_{\rm B}T C 3 (2\pi )^{-5} \int d^3k d\omega \ {1\over \omega} {\rm Im} E(\omega),\eqno(4)$$
where
$$C={8\pi(4-5\nu)\over 45(1-\nu^2)}.$$
In deriving (4) we have used the relation $\mu = E/2(1+\nu)$, where $E(\omega)$ is the Young's
(viscoelastic) modulus.
The shear stress when we only include wavevectors up to $k=2\pi /D$ (where $D$ is a cut-off length, of order
the distance between the cross-links, or larger)
is given by 
$$\langle \sigma_\parallel^2\rangle = - k_{\rm B}T C D^{-3} {2\over \pi} 
\int_0^\infty d\omega \ {1\over \omega} {\rm Im} E(\omega)$$
Using the {\it sum rule}\cite{Ueba,Brener}
$${2\over \pi} \int_0^\infty d\omega \ {1\over \omega} {\rm Im}E(\omega) 
=  E_0 -E_\infty $$
where $E_0=E(0)$ and $E_\infty = E(\infty)$ are the low- and high-frequency
rubber modulus (both real), 
we get
$$\bar \sigma_\parallel^2\equiv \langle \sigma_\parallel^2 \rangle 
= k_{\rm B} T  C D^{-3} (E_\infty - E_0)$$
Assuming $\nu = 0.5$ gives $2\pi C/3 \approx 1.1$.
In a typical case $E_\infty \approx 10^9 \ {\rm Pa}>> E_0$ and if $D=10 \ {\rm nm}$ we get at room temperature
$\bar \sigma_\parallel \approx 
1 \ {\rm MPa}$,
which (typically) is of the same order of magnitude as the depinning stress at a 
rubber-substrate interface.

It is also interesting to estimate the fluctuation in the displacement ${\bf u}_\parallel$
and the fluctuation in the strain.
Using the same approach as above one obtain
$$\langle u_\parallel^2 \rangle =  \langle u_x^2+u_y^2\rangle = k_{\rm B}T C' D^{-1} {2\over \pi}  
\int_0^\infty d\omega \ {1\over \omega} {\rm Im} \left ( {1\over E(\omega)}\right )$$
where
$$C'=  {2(5-6\nu)(1+\nu)\over 3(1-\nu)}$$
If we use the sum rule\cite{Ueba,Brener}
$${2\over \pi} \int_0^\infty d\omega \ {1\over \omega} {\rm Im}\left ({1\over E(\omega)}\right ) 
= {1\over E_0} -{1 \over E_\infty }$$
we get
$$\langle u_\parallel^2 \rangle =   k_{\rm B}T C' D^{-1}  
 \left ({1\over E_0} -{1 \over E_\infty }\right )$$
In a similar way one can calculate the average of the square of the strain
$$\langle \epsilon_\parallel^2 \rangle =  \langle \epsilon_{zx}^2+\epsilon_{zy}^2\rangle
= k_{\rm B}TC'' D^{-3}  
 \left ({1\over E_0} -{1 \over E_\infty }\right )$$
where
$$C''= {32 \pi (4-5\nu)(1+\nu)^2\over 45 (1-\nu^2)}$$
With $E_0=1 \ {\rm MPa}$ and $D=10 \ {\rm nm}$ one get at 
room temperature $\bar u_\parallel \approx 1 \ {\rm nm}$ and $\bar \epsilon_\parallel \approx 0.3$.

The analysis above indicate that, on the length scale of $\sim 10 \ {\rm nm}$, very large
strain and stress fluctuations occur in normal rubber. The stress fluctuations are of similar magnitude
as the (typical) rubber-substrate depinning stress, suggesting that 
thermally excited transitions over the (lateral) 
energy barriers play a crucial
role in rubber friction on smooth substrates. However, the stress fluctuations ($\sim 1 \ {\rm MPa}$)
are negligible compared to the stress necessary for detaching a rubber patch in the normal direction,
since the latter is determined by the adhesional stress\cite{superlub,SSR} 
which typically is of order $1 \ {\rm GPa}$.
Thus, rubber sliding on smooth surfaces never involves (thermally activated) 
detachment of rubber from the substrate surface, but only (lateral)
sliding of rubber patches (the lateral depinning stress is typically $100-1000$ times smaller than 
the adhesional stress). [Detached areas may form and propagate at the interface (as for Schallamach waves,
see fig. \ref{four}(d)), but in these cases they are generated by the external
applied stress, and depend on the shape of the bodies; the detached regions usually form at
the edge of the rubber-substrate contact region, where elastic instabilities (e.g., ``buckling'') 
of the rubber may occur\cite{instability}.]
In the next section we will develop a model of rubber sliding friction based
on the picture presented above.

\vskip 0.5cm
{\bf 3. Theory}

\begin{figure}
\includegraphics[width=0.35\textwidth,angle=0]{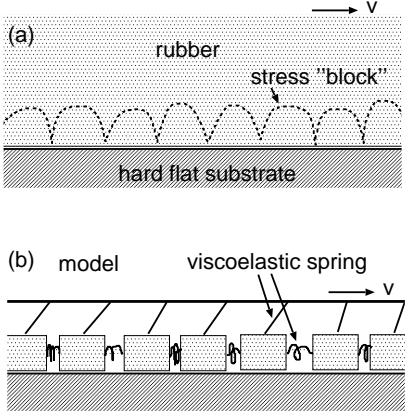}
\caption{\label{model.picture}
(a) Rubber block in adhesive contact with a flat substrate. During sliding
at low velocities, small rubber volume elements (stress blocks) perform stick-slip
motion. (b) The model used in the mathematical description of the system
in (a). The viscoelastic springs $k$ are determined by the rubber viscoelastic
modulus and by the lateral size $D$ of the stress 
blocks via $k(\omega) \approx E(\omega) D$. 
}
\end{figure}

We consider a rubber block with a smooth flat surface sliding on
a perfectly smooth and flat substrate. We assume that the adhesive rubber-substrate interaction will
result in perfect contact between the two solids (i.e., we assume that no Schallamach waves occur
at the interface). During sliding at low velocities, small (nanometer sized) 
regions (stress blocks) at the sliding interface will perform stick-slip motion. We model
the real physical system in Fig. \ref{model.picture}(a) with the block-spring model shown 
in Fig. \ref{model.picture}(b).
However, the springs in the model are not elastic springs but {\it viscoelastic} 
springs determined by the viscoelastic modulus
of the rubber (see below). Furthermore, the blocks in the model experience not only the 
stress from the substrate and the forces from the springs, but also randomly fluctuating
(in time) forces derived from the thermal motion of the molecules in the solid. The strength
of the fluctuating forces is determined by the temperature and by the viscoelastic properties of the springs
via the fluctuation-dissipation theorem. The combination of the (thermal) fluctuating forces, 
and the forces
derived from the external pulling of the upper surface of the rubber block, and the stress acting on the
bottom surface of the block from the substrate, determine the motion of the stress blocks.  

\vskip 0.3cm
{\bf 3.1. Basic equations}

We assume that the motion at low sliding velocity occur by a
thermally activated process, where small surface areas 
or ``patches''
of rubber at the interface perform stick-slip motion. 
We will refer to these ``patches''
as stress domains.
The stress domains are pinned by the substrate potential, and we assume that
some characteristic shear stress $\sigma_c$ must be reached before local slip
can occur. Thus, the pinning force $F_c = \sigma_c D^2$, where $D$ is the
characteristic linear size
of a stress domain. 

The equation of motion for the coordinate $q_i$ of the stress domain $i$
is assumed to be
$$m\ddot q_i = \hat k_z (x-q_i) + \hat k_x (q_{i+1}-q_i)$$
$$+ \hat k_x (q_{i-1}-q_i)+f_i +F_i\eqno(5)$$
where $x=vt$ and where $\hat k_z$ is a time integral operator
$$\hat k_z x(t) = \int_{-\infty}^t dt' \ k_z(t-t') x(t')\eqno(6)$$ 
and similar for $\hat k_x$. We consider $N$ stress domains, $i=1,..., N$,
and assume periodic boundary conditions so that $q_0=q_N$. 
In (5), $f_i(t)$ is a stochastically fluctuating force which result from the
thermal motion of the rubber molecules. 
The force $F_i$ act on the 
stress domain $i$ from the substrate and is defined as follows:
When the stress domain $i$ slips, then $F_i =-m\eta \dot q_i$. 
In the pinned state, $F_i$ is just large enough to balance the total force
exerted by the stress block on the substrate:
$$F_i = -\hat k_z (x-q_i) - \hat k_x (q_{i+1}-q_i)- \hat k_x (q_{i-1}-q_i)-f_i$$
Slip will start when $|F_i|$ 
reaches a critical value $F_c$, either as a result 
of the external
applied force or as a result of a large 
enough thermal fluctuation, or, in general,
as a combination of both these effects.

We define the Fourier transform
$$x(\omega) = {1\over 2 \pi}\int dt \ x(t) e^{i\omega t}\eqno(8)$$
$$x(t) = \int d\omega \ x(\omega ) e^{-i\omega t}\eqno(9)$$
The fluctuating force $f_i(t)$ result from the
thermal motion of the rubber molecules and must satisfy the 
fluctuation-dissipation theorem. 
That is, if we write
$$\hat K_{ij} q_j = \hat k_z q_i - \hat k_x (q_{i+1}-q_i)- 
\hat k_x (q_{i-1}-q_i)\eqno(10)$$ 
then from the theory of Brownian motion (see Appendix A)
$$\langle f_i(\omega ) f_j(\omega' )\rangle = - {k_B T\over \pi \omega} {\rm Im}
K_{ij} (\omega) \delta (\omega +\omega')\eqno(11)$$
where
$$K_{ij}(\omega) = \int_0^\infty dt \ K_{ij}(t) e^{i\omega t}$$
and (for $t>0$):
$$K_{ij}(t) = {1\over 2 \pi} \int d\omega \ K_{ij}(\omega ) e^{-i\omega t}.$$
If we write the elastic modulus as\cite{PerssonRubber}
$$E(\omega) =
E_\infty \left (1- 
\sum_n {h_n \over 1-i\omega \tau_n}\right )$$
we get
$$K_{ij}(\omega) =K_{ij}^\infty-K_{ij}^\infty 
\sum_n {h_n \over 1-i\omega \tau_n},\eqno(12)$$
where  
$$K_{ij}^\infty = 
(k_z^\infty+2k_x^\infty)\delta_{ij}
-k_x^\infty (\delta_{i,j+1}+\delta_{i,j-1})
.$$
Eq. (12) gives
$${\rm Im} K_{ij}(\omega) =
-K_{ij}^\infty \sum_n {h_n \tau_n \omega \over |1-i\omega \tau_n|^2}\eqno(13)$$ 
Substituting (13) in (11) gives
$$\langle f_i(\omega ) f_j(\omega' )\rangle 
={k_BT \over \pi} 
K_{ij}^\infty
\sum_n {h_n \tau_n
\over |1-i\omega \tau_n|^2}\delta (\omega+\omega')\eqno(14)$$ 

We can write
$$-
K_{ij} (\omega) q_j (\omega ) +
k_z(\omega ) x(\omega )$$
$$ = -K_{ij}^\infty q_j(\omega) +
k_z^\infty x(\omega) - \sum_n h_n u_{ni} (\omega)\eqno(15)$$
where
$$u_{ni}(\omega)= {-1 \over 1-i\omega \tau_n} K_{ij}^\infty q_j(\omega)+
{1 \over 1-i\omega \tau_n} k_z^\infty x(\omega )
\eqno(16)$$
or
$$u_{ni}(t)+\tau_n \dot u_{ni} (t)= -K_{ij}^\infty q_j(t)+k_z^\infty x+g_{ni}(t)\eqno(17)$$
Here we have added a 
stochastically fluctuating force on the right hand side of (17) which we can
choose so as to reproduce the fluctuating force $f_i(t)$ in Eq. (5). That is,
if we choose $g_{ni}(t)$ appropriately, we can remove the force $f_i(t)$ in
Eq. (5). To this end we must choose 
$$f_i(\omega) = - \sum_n {h_n g_{nj}(\omega) 
\over 1-i\omega \tau_n}\eqno(18)$$
with
$$g_{nj}(\omega) = N^{-1/2} {\rm Re} \sum_k M_{kn} e^{ikx_j} \xi_{kn}\eqno(19)$$
where
$$M_{kn} = \left ({k_BT \tau_n K_k^\infty \over \pi h_n}\right )^{1/2}\eqno(20)$$
with the $k$-sum over 
$$k={2\pi \over D} {r\over N}, \ \ \ \ \ \ \ \ r=0,1,2,...,N-1$$
Note that
$$K_k^\infty = k_z^\infty +k_x^\infty 2[1-{\rm cos}(kD)]\eqno(21)$$
is the (discrete) Fourier transform of $K_{ij}^\infty$. In (19), $\xi_{kn}$
are complex Gaussian random variables:
$$\xi_{kn} = \zeta_{kn}+i\eta_{kn}\eqno(22)$$
where
$$\langle \zeta_{kn}(\omega) \zeta_{k'n'} (\omega') \rangle
= \delta_{nn'} \delta_{kk'}\delta(\omega+\omega')\eqno(23)$$
$$\langle \eta_{kn}(\omega) \eta_{k'n'} (\omega') \rangle
= \delta_{nn'} \delta_{kk'}\delta(\omega+\omega')\eqno(24)$$
$$\langle \zeta_{kn}(\omega) \eta_{k'n'} (\omega') \rangle = 0\eqno(25)$$
Using (19) and (22) gives
$$g_{nj}(\omega) = N^{-1/2} \sum_k M_{kn} \left [\zeta_{kn}{\rm cos} (kx_j)
-\eta_{kn} {\rm sin} (kx_j)\right ]\eqno(26)$$
Using (23)-(25) this gives
$$\langle g_{nl}(\omega) g_{n'l'}(\omega')\rangle $$
$$ = N^{-1}\sum_k M_{kn}^2
{\rm cos} [k(x_l-x_{l'})] \delta_{nn'}\delta (\omega+\omega')$$
$$=N^{-1}\sum_k M_{kn}^2
e^{ik(x_l-x_{l'})} \delta_{nn'}\delta (\omega+\omega')$$
$$={ k_BT \tau_n \over \pi h_n N} \sum_k K_k^\infty 
e^{ik(x_l-x_{l'})} \delta_{nn'}\delta (\omega+\omega')$$
$$={ k_BT \tau_n \over \pi h_n} K_{ll'}^\infty
\delta_{nn'}\delta (\omega+\omega')\eqno(27)$$
Using 
(18) and (27) gives
$$\langle f_i(\omega)f_j(\omega')\rangle
={k_BT \over \pi} 
K_{ij}^\infty
\sum_n {h_n \tau_n
\over |1-i\omega \tau_n|^2}\delta (\omega+\omega')\eqno(28)$$ 
which agree with (14). 

Let us summarize the basic equations:
$$m\ddot q_i = k_z^\infty (x-q_i) +  k_x^\infty (q_{i+1}+q_{i-1}-2q_i)$$
$$ -\sum_n 
h_n u_{ni} +F_i\eqno(29)$$
where
$$u_{ni}(t)+\tau_n \dot u_{ni} (t)= 
k_z^\infty (x-q_i)$$
$$ +  k_x^\infty (q_{i+1}+q_{i-1}-2q_i) 
+g_{ni}(t)\eqno(30)$$
The random force
$$g_{nj}(\omega) = N^{-1/2} {\rm Re} \sum_k M_{kn} e^{ikx_j} \xi_{kn}(\omega) \eqno(31)$$
where
$$M_{kn} = \left ({k_BT \tau_n K_k^\infty \over \pi h_n}\right )^{1/2}.\eqno(32)$$

\vskip 0.5cm
{\bf 3.2. Numerical implementation}

If $D$ is the lateral size of a stress block, then
the mass of a stress block $m=\rho D^3$. We introduce the spring constants
$k_z^\infty = \alpha_z k^*$ and $k_x^\infty = \alpha_x k^*$ where $k^*= D E_\infty$, and 
where $\alpha_x$ and $\alpha_z$ are
of order unity. We measure time in units of $\tau= (m/k^*)^{1/2}$ and distance in unit of
$l=F_c/k^*$ where $F_c = \sigma_c D^2$ is the stress block pinning force. 
We also measure $u$ in units of $F_c$ and $x$ in units of $l$. In these units we get
$$\ddot q_i = \alpha_z^\infty (x-q_i) + \alpha_x^\infty (q_{i+1}+q_{i-1}-2q_i)$$
$$ -\sum_n 
h_n u_{ni} +F_i/F_c\eqno(33)$$
where
$$u_{ni}(t)+(\tau_n/\tau) \dot u_{ni} (t)= 
\alpha_z (x-q_i)$$
$$ +  \alpha_x (q_{i+1}+q_{i-1}-2q_i) 
+g_{ni}(t)\eqno(34)$$
The random force
$$g_{nj}(\omega) = N^{-1/2} {\rm Re} \sum_k M^*_{kn} e^{ikx_j} \xi_{kn} (\omega)$$
where
$$M^*_{kn} = M_{kn}/F_c = \left ({k_BT \tau_n K_k^\infty \over 2\pi \Delta E k^* h_n}\right )^{1/2},$$
where $\Delta E = k^*l^2/2$.
Note also that
$$\langle \zeta_{kn}(\omega) \zeta_{k'n'} (\omega') \rangle
= \delta_{nn'} \delta_{kk'}\delta(\omega+\omega')$$
gives
$$\langle \zeta_{kn}(t) \zeta_{k'n'} (t') \rangle
= 2\pi \delta_{nn'} \delta_{kk'}\delta(t-t')$$
In order to numerically integrate the equations (33) and (34) 
we discretize time with the step length $\delta t$.
The (fluctuating) force  $\delta g_{ni}$ to be used for each time-step in (34) 
can be written as 
$$\delta g_{nj} = N^{-1/2} {\rm Re} \sum_k M^*_{kn} e^{ikx_j} \int_t^{t+\delta t} dt' \xi_{kn}(t')$$
But if
$$\langle \zeta (t) \zeta (t') \rangle
= 2\pi \delta(t-t')$$
we get
$$\langle (\delta \zeta)^2 \rangle
=\int_t^{t+\delta t} dt' dt'' \langle \zeta(t') \zeta(t'') \rangle =2\pi \delta t$$
Thus we can write $\delta \zeta = (2\pi \delta t)^{1/2} G$ where $G$ is a Gaussian random number with
$\langle G^2 \rangle = 1$.
Thus, we take
$$\delta g_{nj} = N^{-1/2} {\rm Re} \sum_k M^*_{kn} e^{ikx_j} 
(2 \pi \delta t)^{1/2} (G_{kn}^{(1)}+iG_{kn}^{(2)})$$
where $G_{kn}^{(1)}$ and 
$G_{kn}^{(2)}$ are Gaussian random numbers. 
We can also write
$$\delta g_{nj} = N^{-1/2} {\rm Re} \sum_k \bar 
M_{kn} e^{ikx_j} (G_{kn}^{(1)}+iG_{kn}^{(2)})\eqno(35)$$
where
$$\bar M_{kn} = 
\left ({k_BT\over \Delta E} {\tau_n \delta t \over \tau^2} {K_k^\infty \over h_n k^*} \right )^{1/2}.$$

In the calculations presented below we have assumed that a sliding block return to the pinned state
when the shear stress $|\sigma| < \sigma_{c1} = \lambda \sigma_c$, where $\sigma_c=F_c/D^2$ is the
depinning stress and $\lambda < 1$. 
We assume that when
the shear stress has decreased below $\sigma_{c1}$
the block return to the pinned state with the 
probability rate $w$, and 
we use random number to determine when the transition actually takes place.
Thus, if $r$ is a random number uniformly distributed 
in the interval $[0,1]$, then if $|\sigma| < \sigma_{c1}$
we assume 
that the stress block return to the pinned state during the time interval $\delta t$
(the time integration step length) if $w\delta t > r$.
In the simulations below we use 
$w \le 10^{12} \ {\rm s}^{-1}$ 
and $\lambda = 0.1$. We use $\delta \approx 10^{-13} \ {\rm s}$ in most of our simulations so that
the condition $w \delta t << 1$ is satisfied, and the results presented below does not depend on
the time step $\delta t$. The calculation of $\delta g_{nj}$ (Eq. (35)) is conveniently performed
using the Fast Fourier Transform method.

\begin{figure}
\includegraphics[width=0.43\textwidth,angle=0]{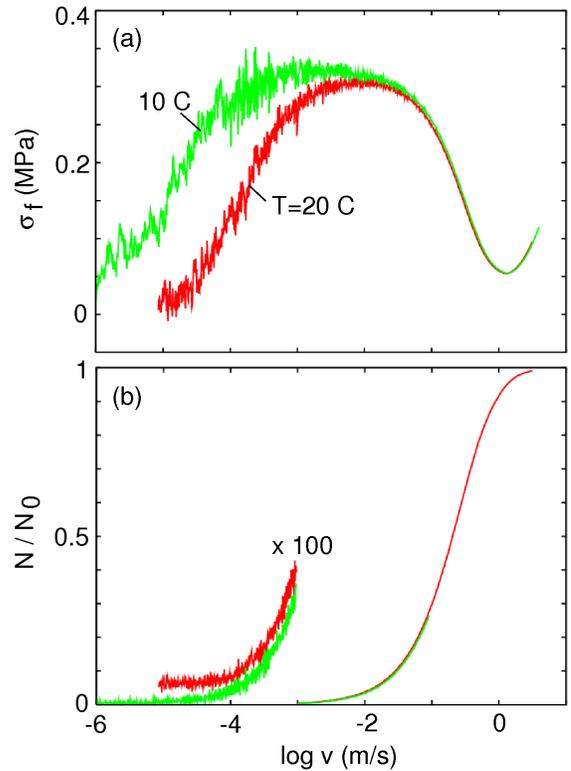}
\caption{\label{sig.N}
(a) The frictional shear stress $\sigma_{\rm f}$ and (b) 
the fraction of slipping surface area $N/N_0$ as a function of the
logarithm (with 10 as basis) of the sliding velocity of the rubber block.
For Styrene butadiene copolymer with 60\% carbon black and for
two different temperatures, $T=10 \ ^\circ {\rm C}$ and $20 \ ^\circ {\rm C}$.
In the calculation we have used the (zero-temperature) depinning stress 
$\sigma_{\rm c} = 1 \ {\rm MPa}$ and the stress block size $D=30 \ {\rm nm}$.
The number of stress blocks was 128. The viscous friction coefficient during
steady sliding $\eta = 0.03$ (natural units) 
and the critical stress below which steady sliding becomes
metastable $\sigma_{\rm c1} = 0.1 \sigma_{\rm c}=0.1 \ {\rm MPa}$. The probability rate per unit time
to return to the pinned state when $\sigma < \sigma_{\rm c1}$ is $w = 2\times 10^{10} \ {\rm s}^{-1}$.
}
\end{figure}

\vskip 0.5cm
{\bf 4. Numerical results}

We now present numerical results for styrene-butadiene rubber both with and
without filler. 

Fig. \ref{sig.N}(a) shows the frictional shear stress $\sigma_{\rm f}$ and (b) 
the fraction of slipping surface area $N/N_0$ as a function of the
logarithm (with 10 as basis) of the sliding velocity of the rubber block.
The results are for styrene butadiene copolymer with 60\% carbon black, and for
two different temperatures, $T=10 \ ^\circ {\rm C}$ and $20 \ ^\circ {\rm C}$.
Note that the low velocity part ($v < v_1$, where $v_1\approx 10^{-2} \ {\rm m/s}$ is the velocity
at which the friction is maximal) of the friction curve is shifted by $\sim 1$ decade
towards lower velocities when the temperature is reduced from $20$ to $10 \ ^\circ {\rm C}$.
This is identical to the change in the (bulk) viscoelastic shift factor $a_T$, 
which change by a factor of $\sim 10$ during the same temperature change.
Fig. \ref{sig.N}(b) shows that the number of moving stress blocks is basically temperature independent
for sliding velocities $v > v_1 \approx 10^{-2} \ {\rm m/s}$, i.e., for $v > v_1$ the fluctuating
force arising from the finite temperature has a negligible influence on the 
friction force. For $v < v_1$ more stress blocks are depinned at the higher temperature
because the fluctuating force is larger at the higher temperature. 
Note also that the maximal shear stress $\sigma_{\rm f} \approx 0.3 \ {\rm MPa}$
is about 30\% of the depinning stress $\sigma_c= 1 \ {\rm MPa}$.

\begin{figure}
\includegraphics[width=0.43\textwidth,angle=0]{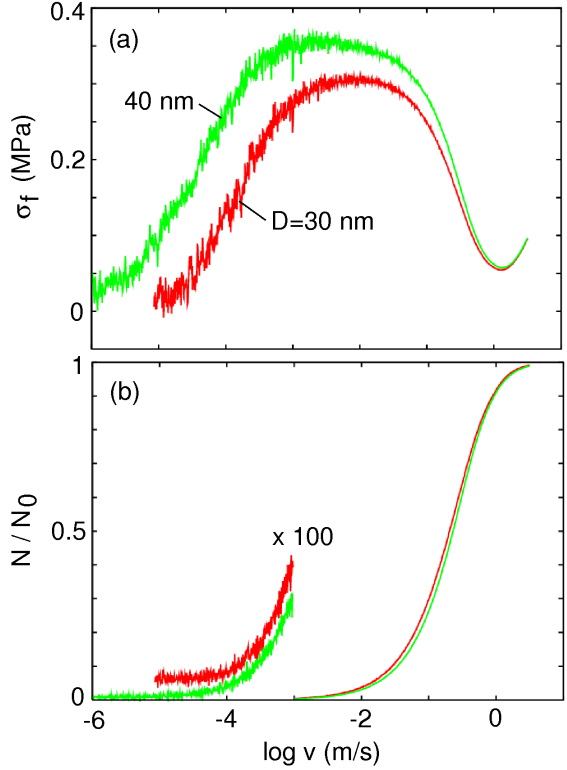}
\caption{\label{sig.D}
The frictional shear stress $\sigma_{\rm f}$  
as a function of the
logarithm of the sliding velocity of the rubber block.
For the temperature
$T=20 \ ^\circ {\rm C}$ and for 
the stress block sizes $D=30$
and $40 \ {\rm nm}$.
All other parameters as in Fig. 5.
}
\end{figure}

Fig. \ref{sig.D}(a) shows the frictional shear stress $\sigma_{\rm f}$  
and (b) the fraction of sliding blocks, 
as a function of the
logarithm of the sliding velocity.
In the calculation we have used $T=20 \ ^\circ{\rm C}$ 
and the stress block sizes $D=30$
and $40 \ {\rm nm}$. 
When the stress block size increases the pinning force $\sigma_c D^2$ increases,
and it is necessary to go to lower sliding velocities in order for temperature effects
to manifest them-self (as a decrease in the frictional shear stress).

The linear size of the stress blocks, $D$, is most likely 
determined by the elastic modulus and the depinning shear stress $\sigma_c$ as follows.
The stress block is the smallest unit which is able to slide as a coherent unit and can
be determined as follows. The depinning force is $F_c=\sigma_c D^2$. If the stress $\sigma$ act 
at the bottom surface of a stress block it will move a distance $x$ determined by $kx = \sigma D^2$
where $k \approx E D$, 
where 
$E$ is the elastic modulus. The stress block experience a quasi-periodic pinning 
potential from the substrate,
characterized by a lattice constant $a$ of order a few Angstroms. Thus, the stress block will in general be able
to occupy a ``good'' binding position in the corrugated substrate potential only if
$ka$ is less than the pinning potential\cite{picture}. The condition $ka=F_c$ gives the size $D$ of the 
pinned domains. Using $F_c=\sigma_cD^2$ and $k \approx E D$ this gives
$$D \approx Ea /\sigma_c.$$
In a typical case $E\approx 10 \ {\rm MPa}$, $\sigma_c \approx 1 \ {\rm MPa}$ and $a$ a few Angstroms,
giving $D = 30 \ {\rm nm}$. Since $D$ increases when the elastic modulus increases, the present theory
indicate that the friction should increase with increasing elastic modulus. This is exactly what has
been observed in friction studies for smooth surfaces (see below). Since the elastic
modulus $E$ depends on frequency, during sliding the size $D$ of the stress domains may depend on
the sliding velocity, 
but we have not taken this effect into account 
in this paper.  

\begin{figure}
\includegraphics[width=0.43\textwidth,angle=0]{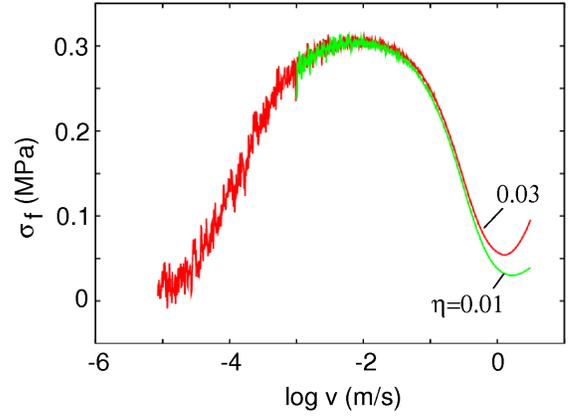}
\caption{\label{sig.eta}
The frictional shear stress $\sigma_{\rm f}$  
as a function of the
logarithm of the sliding velocity of the rubber block.
For the temperature
$T=20 \ ^\circ {\rm C}$ and for 
the viscous friction coefficients 
$\eta = 0.03$ and $0.01$ (natural units).
All other parameters as in Fig. 5.
}
\end{figure}

Fig. \ref{sig.eta} shows the frictional shear stress $\sigma_{\rm f}$  
as a function of the
logarithm of the sliding velocity.
Results are shown for two different viscous friction coefficients
$\eta = 0.03$ and $0.01$ (natural units). 
As expected, $\eta$ is only important at relative high sliding velocities.

\begin{figure}
\includegraphics[width=0.43\textwidth,angle=0]{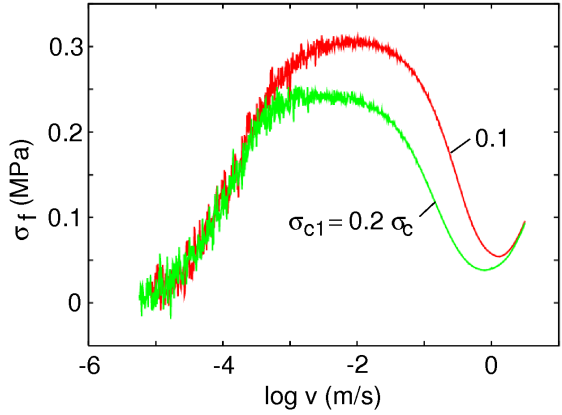}
\caption{\label{sig.sigc1}
The frictional shear stress $\sigma_{\rm f}$  
as a function of the
logarithm of the sliding velocity of the rubber block.
For the temperature
$T=20 \ ^\circ {\rm C}$ and for 
the critical stresses (below which steady sliding becomes
metastable) $\sigma_{\rm c1} = 0.1 \sigma_{\rm c}=0.1 \ {\rm MPa}$ and $0.2 \ {\rm MPa}$. 
All other parameters as in Fig. 5.
}
\end{figure}

Fig. \ref{sig.sigc1} shows how $\sigma_{\rm f}$  
depend on the sliding velocity 
for two different values 
($0.1$ and $0.2 \ {\rm MPa}$) 
of the critical stress $\sigma_{\rm c1}$ below which the 
sliding patch can return to the pinned state.

\begin{figure}
\includegraphics[width=0.43\textwidth,angle=0]{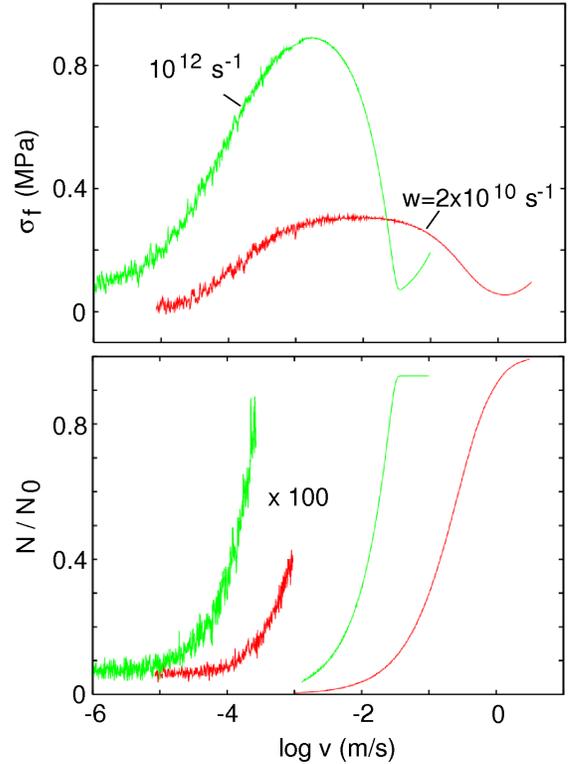}
\caption{\label{varyw}
The frictional shear stress $\sigma_{\rm f}$  
as a function of the
logarithm of the sliding velocity of the rubber block.
For the temperature
$T=20 \ ^\circ {\rm C}$ and for 
the probability rates per unit time
to return to the pinned state (when $\sigma < \sigma_{\rm c1}$) 
$w = 2\times 10^{10} \ {\rm s}^{-1}$ and $w = 10^{12} \ {\rm s}^{-1}$.
All other parameters as in Fig. 5.
}
\end{figure}

\begin{figure}
\includegraphics[width=0.43\textwidth,angle=0]{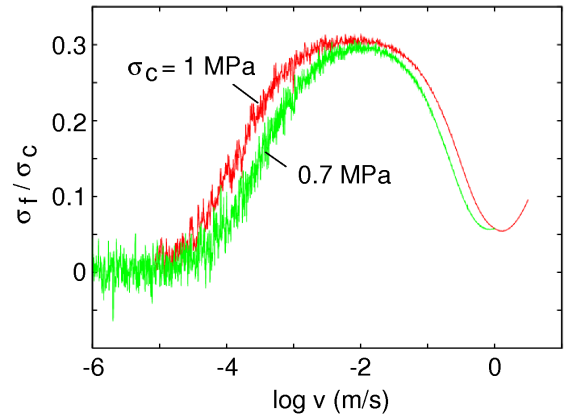}
\caption{\label{varysig}
The frictional shear stress $\sigma_{\rm f}$  
as a function of the
logarithm of the sliding velocity of the rubber block.
For the temperature
$T=20 \ ^\circ {\rm C}$ and for 
the (zero-temperature) depinning stresses 
$\sigma_{\rm c} = 1 \ {\rm MPa}$ and and $\sigma_{\rm c} = 0.7 \ {\rm MPa}$.
All other parameters as in Fig. 5.
}
\end{figure}

Fig. \ref{varyw} shows the frictional shear stress $\sigma_{\rm f}$  
as a function of the
logarithm of the sliding velocity of the rubber block
for two different values 
($2\times 10^{10}$ and $10^{12} \ {\rm s}^{-1}$)
of the probability rate per unit time
to return to the pinned state when $\sigma < \sigma_{\rm c1}$.
Fig. \ref{varysig} shows similar results for two different values 
($1$ and $0.7 \ {\rm MPa}$) 
of the
depinning stress $\sigma_{\rm c}$.

\begin{figure}
\includegraphics[width=0.43\textwidth,angle=0]{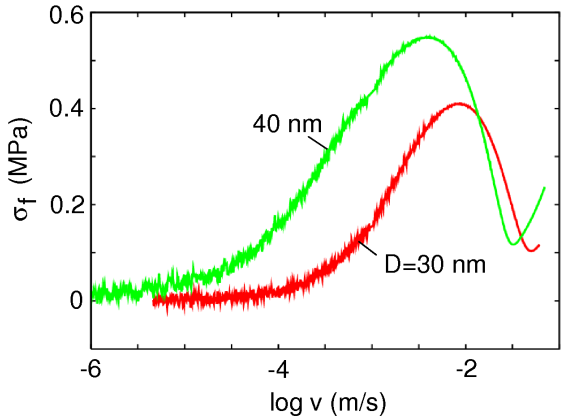}
\caption{\label{unfilled}
The frictional shear stress $\sigma_{\rm f}$  
as a function of the
logarithm of the sliding velocity of the rubber block.
For unfilled styrene butadiene (SB) copolymer 
for
$T=20 \ ^\circ {\rm C}$.
In the calculation we have used the (zero-temperature) depinning stress 
$\sigma_{\rm c} = 1 \ {\rm MPa}$ and the stress block 
size $D=30 \ {\rm nm}$ and $40 \ {\rm nm}$.
The number of stress blocks was 128. The viscous friction coefficient during
steady sliding $\eta = 0.03$ (natural units), 
and the critical stress below which steady sliding becomes
metastable $\sigma_{\rm c1} = 0.1 \sigma_{\rm c}=0.1 \ {\rm MPa}$.
The probability rate per unit time
to return to the pinned state when $\sigma < \sigma_{\rm c1}$ is $w = 10^{12} \ {\rm s}^{-1}$.
}
\end{figure}

Fig \ref{unfilled} shows the frictional shear stress $\sigma_{\rm f}$  
as a function of the
logarithm of the sliding velocity of the rubber block for an
unfilled styrene butadiene (SB) copolymer for two different values 
($30$ and $40 \ {\rm nm}$) of the size $D$ of a pinned region.

\vskip 0.5cm
{\bf 5. Discussion}

Vorvolakos and Chaudhury\cite{Cha} have performed a very detailed experimental study of 
sliding friction for
silicon rubber on hard flat (passivated) substrates (see also Ref. \cite{Grosch,Reiter,Caroli} for
other studies of elastomer friction).
They used silicon rubbers with many different (low-frequency) elastic modulus $E$.

\begin{figure}
\includegraphics[width=0.45\textwidth,angle=0]{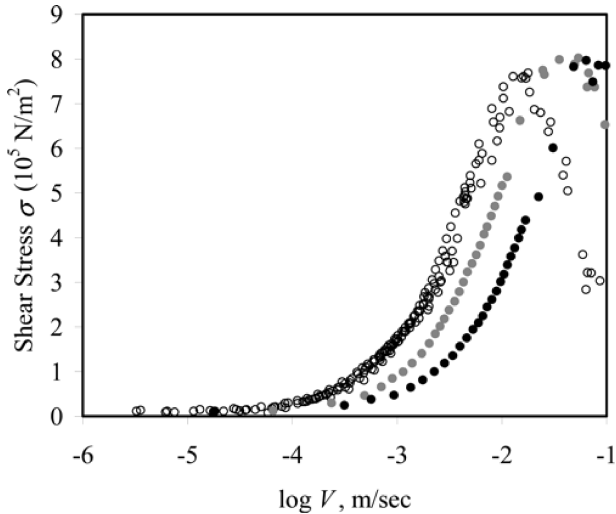}
\caption{\label{L4}
Shear stress as a function of velocity and temperature for
a silicon elastomer (low frequency elastic modulus $E\approx 5 \ {\rm MPa}$)
sliding on a Si wafer covered by an inert self-assembled monolayers
film. Open circles, gray circles and black
circles represent data at 298, 318 and $348 \ {\rm K}$, respectively. Adapted from
Ref. \cite{Cha}
}
\end{figure}

In Fig. \ref{L4} we show the velocity dependence of the shear stress as measured at different temperatures
$T=298$ (open circles), 318 (gray circles) and $348 \ {\rm K}$ (black circles) for a silicon rubber
with the low-frequency elastic modulus $E\approx 5 \ {\rm MPa}$.
The experimental data on the low-velocity side can be shifted to a 
single curve when plotted as a function
of $va_T$, see Ref. \cite{Cha}. This is in accordance with our model calculations (see Sec. 4) and shows the
direct involvement of the rubber bulk in the sliding dynamics. 

Vorvolakos and Chaudhury\cite{Cha} have shown that 
the frictional shear stress for all the studied rubbers
increases with increasing elastic modulus.
This is in qualitative agreement with our theory since, as explained in Sec. 4,
as $E$ increases we expect the linear size $D$ of the stress domains to increase,
which will increase the sliding friction (see Fig. \ref{sig.D} and \ref{unfilled}).

In the experiments by Grosch\cite{Grosch} it was observed that the friction on smooth surfaces 
has a bell-like shape with a maximum at some characteristic velocity $v=v_1$. 
We observe the same general behavior, but with some important differences.
Thus, the experimental data of Grosch was observed to obey the WLF transform. That is, the full
friction curve could be constructed by performing measurements in a very limited velocity
range and at different temperatures, and then use the WLF transform to shift the 
measured data to a single temperature. We also find that our calculated friction
obey the WLF transform for $v < v_1$, but not for $v>v_1$. Thus, the decrease in the
friction which we observe for $v>v_1$
is nearly temperature independent. 
The difference between our prediction and the Grosch results can be understood as follows:
Most likely, the decrease of the friction for $v>v_1$ in the Grosch experiment is related to
a decrease in the area of real contact with increasing sliding velocity. 
The Grosch experiments where performed on smooth but wavy glass, and 
the area of real contact depends on the effective elastic modulus of the rubber. Thus, as the sliding
velocity increases or, equivalently, the temperature decreases,
the rubber becomes more stiff and the area of real contact decreases. Hence,
if the shear stress remains constant at high sliding velocity 
[as we indeed observe in our calculations if the calculations are 
performed at {\it low velocities} and 
different temperatures, 
and shifted to higher velocity using the WLF equation], then the friction force will decrease
with increasing sliding velocity because of the decrease in the area of real contact. 

\begin{figure}
\includegraphics[width=0.43\textwidth,angle=0]{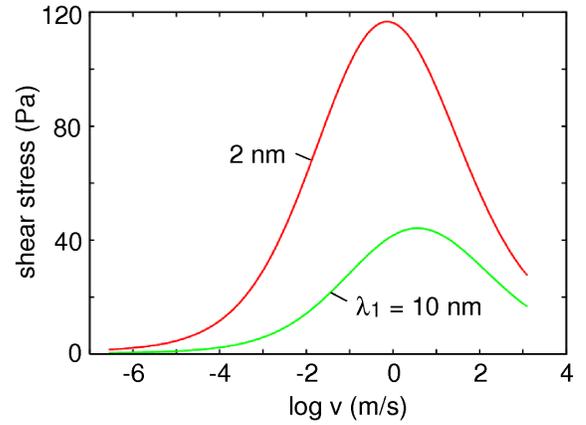}
\caption{\label{shear}
The frictional shear stress $\sigma_{\rm f}$  
as a function of the
logarithm of the sliding velocity of the rubber block.
For Styrene butadiene copolymer with 60\% carbon black and for
$T=40 \ ^\circ {\rm C}$.
The friction is entirely due to the surface roughness of the substrate
which is assumed to be self affine fractal with the fractal dimension $D_{\rm f}=2.3$
and the root-mean-square roughness $0.5 \ {\rm nm}$. The long-wavelength and short wavelength
roll-off wavevectors
$q_0 = 2\pi /\lambda_0$ and $q_1=2\pi /\lambda_1$ where $\lambda_0 = 100 \ {\rm \mu m}$ and
$\lambda_1=2 \ {\rm nm}$ (upper curve) and $10 \ {\rm nm}$ (lower curve).
}
\end{figure}

\begin{figure}
\includegraphics[width=0.43\textwidth,angle=0]{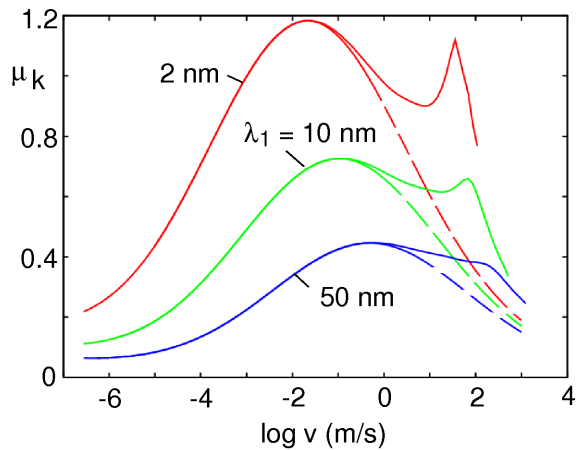}
\caption{\label{Muk}
The kinetic friction coefficient $\mu_{\rm k}$  
as a function of the
logarithm (with 10 as basis) of the sliding velocity of the rubber block.
For Styrene butadiene copolymer with 60\% carbon black and for
$T=40 \ ^\circ {\rm C}$.
The friction is entirely due to the surface roughness of the substrate
which is assumed to be self affine fractal with the fractal dimension $D_{\rm f}=2.3$
and the root-mean-square roughness $1 \ {\rm \mu m}$. 
The long-wavelength and short wavelength
roll-off wavevectors
$q_0 = 2\pi /\lambda_0$ and $q_1=2\pi /\lambda_1$ where $\lambda_0 = 100 \ {\rm \mu m}$ and
$\lambda_1=2 \ {\rm nm}$ (top curve), $10 \ {\rm nm}$ (middle curve) and 
$50 \ {\rm nm}$ (bottom curve). The solid curves are the full calculation while the dashed curves 
are without the flash temperature effect.
}
\end{figure}

\vskip 0.5cm
{\bf 6. The role of surface roughness: application to rubber sealing}

The experiments by Vorvolakos and Chaudhury\cite{Cha} was performed on (passivated) silicon wafers
with the root-mean-square roughness of at most $0.5 \ {\rm nm}$. For such smooth surfaces
the adhesive rubber-substrate interaction will, even in the absence of
a squeezing pressure, give rise to complete contact between the 
rubber and the substrate within the nominal contact area. For this case we have applied the rubber
friction theory developed in Ref. \cite{P8} to obtain the contribution to the frictional
shear stress from the roughness induced bulk viscoelastic 
deformations in the rubber (Fig. \ref{four}c). We have
assumed that the substrate is self affine fractal with the fractal dimension $D_{\rm f} = 2.3$
and the long-distance roll off wavelength $\lambda_0 = 100 \ {\rm \mu m}$. We have included roughness
components down to the short-distance (cut-off) wavelength $\lambda_1$. In Fig. \ref{shear} we show the resulting
frictional shear stress (assuming perfect contact at the sliding interface) using $\lambda_1 = 10 \ {\rm nm}$
and $2 \ {\rm nm}$. The smallest possible $\lambda_1$ is determined by an atomic distance but a more likely
cut-off length in the present case is the mean distance between rubber cross links, which is of order
a few nanometer.
The calculated maximal shear stress is of order $\sim 100 \ {\rm Pa}$, which is a factor
$\sim 10^4$ smaller than the shear stress observed 
in Ref. \cite{Cha}. Hence we conclude that the surface roughness
in the measurements of Vorvolakos and Chaudhury has a negligible 
influence on the observed frictional shear stress. However, the surfaces used in 
Ref. \cite{Cha} are exceptionally smooth, and it is of great interest to study the 
influence of the surface roughness on rubber friction for surfaces of
more common use in rubber applications.

One very important application of rubber is for seals.
In a typical such application the rubber is sliding on a (lubricated) steel surface with
the root-mean-square roughness of order $1 \ {\rm \mu m}$. At low sliding velocity (e.g., at the start of
sliding) the lubricant fluid is squeezed away from the rubber-substrate 
asperity contact regions, giving friction
coefficients similar to those of dry (unlubricated) surfaces, typically of order $\mu_{\rm k} \sim 0.5$.
 
\begin{figure}
\includegraphics[width=0.43\textwidth,angle=0]{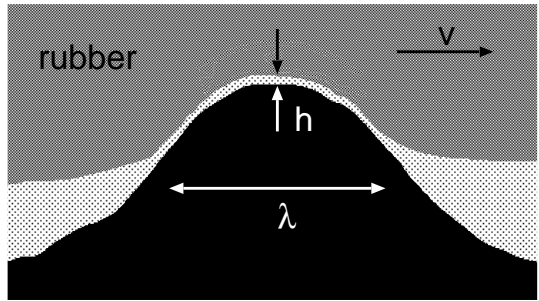}
\caption{\label{filmthickness}
Rubber in squeezed contact with a hard rough substrate. During sliding the bulk
viscoelastic deformation (on the length scale $\lambda$) 
at an asperity contact region will not be influenced by the thin
lubrication film (thickness $h$) at the asperity if $h << \lambda$. The lubrication film will,
however, remove the contribution to the friction from substrate 
asperities with height smaller than $h$.
}
\end{figure}

Recently Mofidi et al\cite{Mofi} have performed rubber friction 
studies for nitrile rubber (the most popular seal 
material) in contact with a steel surface.
The experiments
was performed by squeezing a steel cylinder (diameter $1.5 \ {\rm cm}$
and length $2.2 \ {\rm cm}$) with the force $F_{\rm N} = 100 \ {\rm N}$
against the nitrile rubber surface,
lubricated by 
various types of oil with kinetic viscosities ranging from $\nu = 5.5 \times 10^{-6}$ to
$46 \times 10^{-6} \ {\rm m^2/s}$. 
The nominal squeezing pressure $p \approx 1 \ {\rm MPa}$.
The steel cylinder
was oscillating along its axis with the frequency $f=50 \ {\rm Hz}$ and the amplitude $ a = 1 \ {\rm mm}$.
Thus, the average sliding velocity $\bar v \approx \omega a /\surd 2 \approx 0.2 \ {\rm m/s}$
(where $\omega = 2 \pi f$ is the angular frequency). The steel surface had the rms 
roughness $1.03 \ {\rm \mu m}$. The experiment showed that the kinetic friction $\mu_{\rm k} \approx 0.6$ 
was nearly independent of
the lubrication oil viscosity. This indicate that the lubrication oils are nearly completely removed from the 
steel-rubber asperity contact regions and that the observed friction is due to the asperity induced 
deformation of the rubber surface\cite{P8}. For perfectly smooth surfaces,
the thickness $h$ of the oil film in the  
contact region is easy
to estimate using the equation\cite{PerssonBook,Mugele}
$h \approx (\rho \nu A /p t)^{1/2}$, where $\rho$ is the oil
mass density, $\nu$ the kinematic viscosity, $A$ the nominal contact area, and $t$ the squeezing time
which in the study presented in Ref. \cite{Mofi} was $15 \ {\rm minutes}$. In the present case this 
gives $h \approx 20-60 \ {\rm nm}$. The actual film thickness at the asperity contact 
regions (see Fig. \ref{filmthickness}) may be much smaller.

We now show that the rubber friction theory developed in Ref. \cite{P8}, where the whole friction is
attributed to the asperity-induced viscoelastic deformations of the rubber, can in fact explain the
magnitude of the friction observed by Mofidi et al\cite{Mofi} (which is similar to the friction observed for
rubber seals during start-up of sliding\cite{KGK}). 
Since the sliding surfaces are embedded in oil we will assume that there is negligible
adhesive interaction between the surfaces so that the area of real contact is determined
mainly by the applied pressure. 
In Fig. \ref{Muk} we show 
the calculated kinetic friction coefficient $\mu_{\rm k}$
as a function of the
logarithm of the sliding velocity of the rubber block.
In the calculation we have assumed styrene butadiene (SB) copolymer with 60\% carbon black and 
$T=40 \ ^\circ {\rm C}$.
The friction is entirely due to the surface roughness of the substrate
which is assumed to be self affine fractal with the fractal dimension $D_{\rm f}=2.3$
and the root-mean-square roughness $1 \ {\rm \mu m}$.
The long-wavelength
roll-off wavevector
$q_0 = 2\pi /\lambda_0$ 
with $\lambda_0 = 100 \ {\rm \mu m}$, and the short wavelength cut-off wavevector
$q_1=2\pi /\lambda_1$ where 
$\lambda_1=2 \ {\rm nm}$ (top curve), $10 \ {\rm nm}$ (middle curve) and
$50 \ {\rm nm}$ (bottom curve). The solid curves are the full calculation while the dashed curves
are without the flash temperature effect\cite{P8,PerssonRubber}. For the $\lambda_1=50 \ {\rm nm}$ case,
according to the calculation the area of real contact between the rubber and the substrate
is only of order $0.3\%$ of the nominal contact area so the frictional shear stress 
(at the shortest scale $\sim 50 \ {\rm nm}$) is of order $\mu_{\rm k} p /0.003 \approx 100 \ {\rm MPa}$
which is close to the (ideal) rupture stress of rubber at 
short length scale (the rupture stress at the
macroscopic scale is usually much smaller due to ``large'' 
crack-like defects in most rubber objects). Thus, one may expect 
rubber wear to occur when sliding under the
present conditions, as was also observed in the study reported on in Ref. \cite{Mofi}. 

In the context of rubber sealing, the surfaces of the hard countermaterials (usually steel)
are prepared to have surface roughness with a rms amplitude
in the range $\sim 0.1-1 \ {\rm \mu m}$, which will trap lubrication oil at the interface.
After a long time of stationary contact lubricant fluid will only occur in the substrate
valleys, but at the onset of sliding lubrication fluid is dragged
out from the valleys (or cavities) to form (at high enough sliding velocity) a thin lubrication film
between the substrate asperities and the rubber surface; this will reduce the friction and the 
rubber wear. 

The theories presented above predict that the rubber friction decreases for 
high enough sliding velocity. This may result in stick-slip motion
(as often observed, e.g., for rubber wiper blades), and is a major problem in many practical
applications as it may generate 
strong noise and wear, and may lead to device malfunction.

\begin{figure}
\includegraphics[width=0.35\textwidth,angle=0]{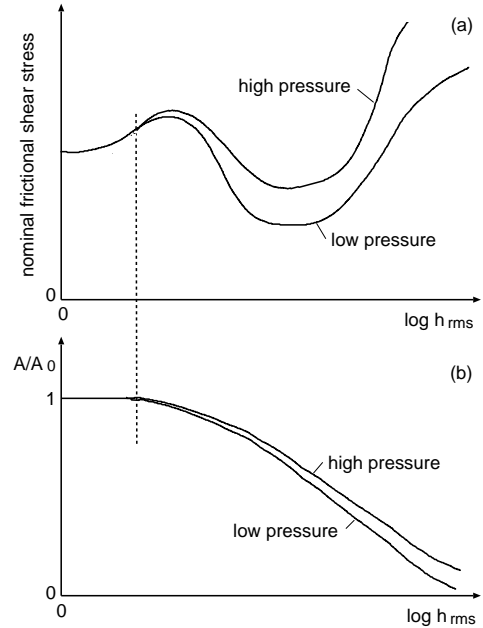}
\caption{\label{qualitative}
The nominal frictional shear stress ($F_{\rm f}/A_0$) (a) and 
the projected (on the $xy$-plane) rubber-substrate
contact area $A$ (normalized by the nominal contact area $A_0$) (b)
during sliding of a rubber block on substrates
with increasing root-mean-square roughness $h_{\rm rms}$.
(Schematic.)
}
\end{figure}

We note, as a curiosity, that the increase in the temperature in the asperity 
contact regions (flash temperature) 
increases the friction at high sliding velocity, see Fig. \ref{Muk}.
This is opposite to some other rubber applications, e.g. for tires, where the flash temperature decreases the
friction\cite{P8,PerssonRubber} This difference 
is due to the fact that in the present case the temperature effect becomes important only at relative
high sliding velocities, where the asperity-induced
perturbing frequencies occur on the glassy-side of the viscoelastic
loss peak (the ${\rm tan}\delta$-peak); as the temperature increases the rubber viscoelastic
spectra shift towards lower frequencies, i.e., the perturbing frequencies will occur closer to
the maximum of the ${\rm tan}\delta$-curve, 
and the friction will increase.

We note that many earlier studies of the adhesional contribution to rubber friction
have used ``polished'' (steel) surfaces\cite{Grosch}. But such surfaces typically have surface
roughness on many different length scales with a rms amplitude of order $1 \ {\rm \mu m}$
and cannot be considered as smooth with respect to rubber friction. In fact, 
as discussed above, the dominant contribution
to the friction in these cases may be derived from the 
substrate asperity induced (viscoelastic) deformation of the rubber
surface.

Finally, let us discuss (qualitatively) the magnitude of rubber friction as a function of
increasing surface roughness. At small surface roughness, because of the rubber-substrate
adhesive interaction, there will be complete contact between the rubber and the substrate
and the rubber friction will be the sum of the contributions illustrated in
Fig. 2 (a) and (c), and the friction will increase with the surface roughness amplitude.
As the roughness increases above some critical value, incomplete contact will occur at the
sliding interface and when the surface roughness becomes large enough the friction may decrease.
However at large enough roughness the rubber friction may increase again, at least if the 
(perpendicular) squeezing pressure is high enough. It is clear that even in the absence of
a lubricant, for sealing-applications the best substrate surface may have some (small)
roughness in order to minimize the friction. 

\vskip 0.5cm
{\bf 7. Summary and conclusion}

We have studied the sliding friction for viscoelastic solids, e.g., rubber,
on hard flat substrate surfaces. We have shown that 
the fluctuating shear stress,
which result from the thermal motion of the atoms 
or molecules in a viscoelastic solid, gives rise to very strong stress-fluctuations, which at
the nanoscale are in the MPa-range. This is similar to the depinning stresses
which typically occur at solid-rubber interfaces, indicating the crucial importance of
thermal fluctuations for rubber friction on smooth surfaces. We have developed a detailed model which
takes into account the influence of thermal fluctuations on the depinning of small 
contact patches (stress domains) at the rubber-substrate interface. The theory predict
that the velocity dependence of the macroscopic shear stress has a bell-shaped form, and that the
low-velocity side exhibit the same temperature dependence as the bulk viscoelastic modulus,
in qualitative agreement with experimental data. Finally, we have discuss the influence
of small-amplitude substrate roughness on rubber sliding friction and shown that in
typical applications to rubber sealing, the substrate asperity-induced viscoelastic
deformations of the rubber surface may give the dominant contribution to
the sliding friction. 

\vskip 0.5cm
{\bf Acknowlengement}:

A.V. and B.P. thank DFG and Pirelli Pneumatici for support. A.V. and B.P. 
thank the EU for support within 
the ``Natribo'' network of the European Science Foundation.  

\vskip 0.5cm
{\bf Appendix A: Memory friction and fluctuating force}

Eq. (11) is a standard result in the general theory of Brownian motion which can
be derived in various ways.
The most general proof is based on the memory function formalism as described, e.g., in the book
by D. Forster\cite{Forster}. A simpler (but less general)
derivation of Eq. (11) involves the study of a particle
(coordinate $q(t)$) coupled to an infinite set of harmonic oscillators 
(the heat bath) (coordinates $x_\mu$). For the readers convenience, we briefly review this
derivation here. The particle and the heat bath coordinates satisfies the equations of motion 
$$m\ddot q +\sum_\mu \alpha_\mu x_\mu=0\eqno(A1)$$
$$m_\mu \ddot x_\mu +m_\mu \omega_\mu^2 x_\mu+\alpha_\mu q +m_\mu \eta_\mu \dot x_\mu = f_\mu\eqno(A2)$$
where 
$$\langle f_\mu (t)f_\nu(t')\rangle = 2 m_\mu \eta_\mu k_BT \delta (t-t')\delta_{\mu \nu}\eqno(A3)$$
If we define
$$x_\mu(t) = \int d\omega \ x_\mu (\omega) e^{-i\omega t}$$
$$x_\mu(\omega ) = {1\over 2 \pi} \int dt \ x_\mu (t) e^{i\omega t}$$
and similar for $q$ and $f_\mu$, we get from (A2)
$$x_\mu(\omega) = {f_\mu(\omega)-\alpha_\mu q(\omega) \over m_\mu (\omega_\mu^2-\omega^2-
i\omega \eta_\mu )}\eqno(A4)$$
and from (A3),
$$\langle f_\mu(\omega )f_\nu(\omega' )\rangle =  m_\mu \eta_\mu k_BT \delta (\omega+\omega')
\delta_{\mu \nu}/\pi. \eqno(A5)$$
From (A4) we get
$$\sum_\mu \alpha_\mu x_\mu(\omega) = 
\gamma(\omega) q(\omega)-f(\omega)\eqno(A6)$$
where
$$\gamma (\omega) =\sum_\mu {\alpha^2_\mu \over m_\mu (\omega_\mu^2-\omega^2-i\omega \eta_\mu )}\eqno(A7)$$
and
$$f (\omega) = \sum_\mu {\alpha_\mu f_\mu(\omega)\over m_\mu 
(\omega_\mu^2-\omega^2-i\omega \eta_\mu )}\eqno(A8)$$
Using (A1) and A(6) gives 
$$-m\omega^2 q(\omega) +\gamma(\omega) q(\omega) = f(\omega)\eqno(A9)$$
Using (A8), (A7) and (A5) it is easy to show that 
$$\langle f(\omega) f(\omega')\rangle = 
- {k_BT\over \pi \omega} {\rm Im} \gamma (\omega) \delta (\omega+\omega')\eqno(A10)$$
Note also that (A9) is equivalent to
$$m\ddot q +\int_{-\infty}^t dt' \ \gamma(t-t') q(t') = f(t)$$
In Sec. 3.1 we studied a system of many coupled dynamical variables
$$m\ddot q_i +\int_{-\infty}^t dt' \ \gamma_{ij}(t-t') q_j(t') = f_i(t)$$
but this problem can be reduced to the problem studied above by forming new dynamical
variables, as linear combination of the old dynamical variables $q_i$, chosen so that 
$\gamma_{ij}$ becomes diagonal.


\begin{thebibliography}{999}
\bibitem{picture}
The force acting on a stress domain should depend quasi-periodically on the
lateral displacement $x$ i.e., $F=F_c{\rm sin}(2\pi x/a)$ giving for small $x$:
$F\approx F_{\rm c} 2 \pi x/a$. The pinning stress $F_{\rm c} = \sigma_{\rm c}D^2$
and the elastic force $F=kx$ with $k=\alpha ED$ where
$\alpha$ is of order $\sim 6$. Thus we get $ED \approx \sigma_{\rm c} D^2/a$.
\end{thebibliography}
\end{document}